# DOUBLE-SLIT INTERFEROMETER MEASUREMENTS AT SPEAR3*

C. L. Li[†], East China University of Science and Technology, Shanghai, China
Y. H. Xu, Donghua University, Shanghai, China
M.J. Boland, Synchrotron Light Source Australia, Clayton, Australia
T. Mitsuhashi, KEK, Tsukuba, Japan
W.J. Zhang, East China University of Science and Technology, Shanghai, China
and the University of Saskatchewan, Saskatoon, Canada
M. Grinberg and J. Corbett, SLAC National Accelerator Laboratory, Menlo Park, USA

## Abstract

The resolution of a conventional telescope used to image visible-light synchrotron radiation is often limited by diffraction effects. To improve resolution, the double-slit interferometer method was developed at KEK and has since become popular around the world. Based on the Van Cittert-Zernike theorem relating transverse source profile to transverse spatial coherence, the particle beam size can be inferred by recording fringe contrast as a function of interferometer slit separation. In this paper, we describe the SPEAR3 double-slit interferometer, develop a theoretical framework for the interferometer and provide experimental results. Of note the double-slit system is 'rotated' about the beam axis to map the dependence of photon beam coherence on angle.

## INTRODUCTION

The synchrotron radiation (SR) interferometer was first developed at KEK by Mitsuhashi [1] and has been widely used since at many facilities worldwide [2]. Compared with an optical telescope, the double-slit interferometer has the advantage of improved spatial resolution. One famous example was Michelson's stellar interferometer [3]. In 1920, Michelson measured the angular diameter of α-Orionis using four mirrors to direct light from the star onto the main telescope at Mt. Wilson to produce interference fringes. The zero order fringes vanished when the outer mirrors were separated by d=3.0734 m. Michelson's measurement gave an estimated angular star diameter of $\varphi$ = 0.047" as calculated from $d \cdot \varphi = \frac{\lambda}{2}$ where λ is the wavelength of the detected light. Based on the known distance of α-Orionis to the telescope, the stellar diameter was determined to be 240x10$^6$ miles assuming a uniform luminous disk (~300x the size of our sun). For Michelson's interferometer, the ratio between the star diameter and the distance to the telescope was about $1 \times 10^{-7}$. For SPEAR3, the ratio is about $1$-$7 \times 10^{-6}$ depending on the electron beam cross-section.

In this paper we describe the SPEAR3 visible light synchrotron radiation (SR) interferometer, provide a theoretical basis for the device and present beam size measurements including data obtained with axial rotation of the double slits.

___________________
* Work supported by US Department of Energy Contract
  DE-AC03-76SF00515, Office of Basic Energy Sciences
[†] email address: chunlei520@gmail.com

## EXPERIMENTAL CONFIGURATION

The double-slit interferometer at SPEAR3 is installed on a visible-light diagnostic beamline. Figure 1 shows a schematic of the beamline with the interferometer optics. Synchrotron radiation from a dipole magnet is first reflected from a Rh-coated pick-off mirror and then transported to the optical bench. A photograph of the unfocused 60mm x 100mm beam at the double slit location is shown in Fig. 1.

For the experiments reported here wavefront division is accomplished using precision pairs of 2mm x 2mm slits cut in a set of cardboard masks. The two resulting beamlets are focused by an f=+1.2m convex mirror and re-imaged by an f=+24mm Takahashi LE24 eyepiece to magnify the interference pattern. A Glan-Thompson polarizer inserted with transmission axis orientated along the horizontal axis selects the σ-mode component of the SR and a Point Grey FL2-08S2M CCD camera captures the interference image for display and analysis in Matlab. Both the double-slit mask and the CCD camera can be mounted on rotatable stages to enable interferometric measurements at arbitrary axial angles relative to the beam axis.

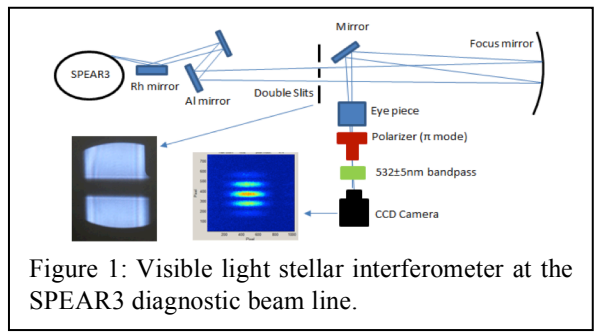

Figure 1: Visible light stellar interferometer at the SPEAR3 diagnostic beam line.

To measure beam size in the horizontal direction, the slit separation was scanned from 8mm to 17.88 mm using two manual translation stages with a measurement step size of about 0.5 mm. In the vertical direction, slit separation was ranged from 25 mm to 85 mm with a step size of 5 mm using a set of discrete insertable masks. To characterize angular dependence, a 16.94 mm fixed-separation double slit system was rotated from 0 to 180 degrees with a step size of 5 degrees.

## THEORETICAL FRAMEWORK

In this section we present a theoretical framework for the experimental measurements. Starting with the expression for a double-slit interferogram produced by

incoherent light of equal intensity $I_o$ on two slits (no diffraction effects), the fringe pattern is [4, 5]

$$I(x,d) = 2I_o(1 + \gamma(d)\cos(x,d)) \qquad (1)$$

where x is position in the interference pattern, d is the slit separation and γ is the normalized cross-correlation factor which characterizes the contrast or visibility in the interference pattern (degree of coherence) [4, 5]. A single, monochoromatic plane-wave, for instance, produces the classic full-contrast γ=1 interference pattern for all slit separations d.

For a distributed source, the electromagnetic field can be modeled as a linear combination of plane waves. The differential path length from each elemental radiator dσ at the source to interferometer slits 1 and 2 varies with position of the element. As a result, each elemental fringe pattern has a different relative phase shift and the net interference pattern suffers loss of contrast, γ<1. This observation was the basis for Michelson's famous measurement of stellar diameters and for Mitsuhashi's application to monitor SR sources.

Using a 'distant source' approximation, the characteristic contrast of the interference pattern is a linear distribution-weighted summation over all elemental sources. In one dimension, along transverse coordinate x, the resulting fringe contrast is the inverse Fourier transform of the normalized source intensity distribution $I(x)$:

$$\gamma(f_x) = \int I(x)e^{+i\vec{k}\cdot\vec{x}}dx \qquad (2)$$

where $L$ is the distance from source to slits, $\vec{k}$ is the wavevector and $f_x = \frac{d}{\lambda L}$ $m^{-1}$ is the spatial frequency [6].

Equation 2 is a simplified version of the Van Cittert-Zernike theorem relating incoherent source intensity to interference fringe contrast [4, 5]. Evaluating the phase term in the exponent we have

$$\vec{k}\cdot\vec{x} \approx \frac{2\pi}{\lambda}\cdot x \cdot \frac{d}{L} \triangleq 2\pi f_x x$$

where $\frac{d}{L}$ is the directional cosine and $f_x$ is again the spatial frequency. For the SPEAR3 interferometer $f_x \sim 10^3$-$10^4$ for green light.

If we assume a 1-D Gaussian source the normalized degree of coherence in the interference pattern is

$$\gamma(f_x) = I_o \int e^{-\frac{x^2}{2\sigma_x^2}} e^{-i2\pi f_x x} dx \qquad (3a)$$

$$\gamma(f_x) = e^{-2f_x^2 \pi^2 \sigma_x^2} \qquad (3b)$$

where $\sigma_x$ is the rms source size.

From Eq. 3b, a single fringe contrast measurement $\gamma_d$ at fixed slit separation d (spatial frequency $f_x$) yields

$$\sigma_x = \frac{1}{\pi f_x}\cdot\sqrt{\frac{1}{2}\ln\left(\frac{1}{\gamma_d}\right)} \quad \text{or} \quad \sigma_x = \frac{\lambda L}{\pi d}\cdot\sqrt{\frac{1}{2}\ln\left(\frac{1}{\gamma_d}\right)} \qquad (4)$$

This deterministic result enables fast on-line beam size measurements with a fixed double-slit separation d.

Similarly, by scanning $f_x$ (slit separation d) through a range of values the measured fringe contrast dependence can be numerically fit to a Gaussian profile yielding

$$\gamma(f_x) = e^{-\frac{f_x^2}{2\sigma_\gamma^2}} \qquad (5)$$

where $\sigma_\gamma$ is the rms width of the contrast curve. Comparing Eq. 5 to Eq. 3b yields the identity

$$\sigma_\gamma^2 = \frac{1}{4\pi^2\sigma_x^2} \quad \text{or} \quad \sigma_\gamma\sigma_x = \frac{1}{2\pi} \qquad (6)$$

reminiscent of transform-limited pulse analysis or the photon beam brightness calculation for a pure Gaussian mode. Expressing $\sigma_\gamma$ in terms of slit separation $\sigma_\gamma(d)$ rather than spatial frequency yields the beam size formula $\sigma_x = \frac{\lambda L}{2\pi\sigma_\gamma(d)}$.

In two dimensions the Van Cittert-Zernike theorem can be written

$$\gamma(f_x, f_y) = \iint I(x,y)e^{+i2\pi(f_x x + f_y y)}dxdy \qquad (7)$$

where the configuration space axes (x, y) and spatial frequencies ($f_x$, $f_y$) are referenced to an x-y Cartesian coordinate system. From Eq. 7 a bi-Gaussian beam profile

$$I(x,y) = I_o e^{-\left(\frac{x^2}{2\sigma_x^2} + \frac{y^2}{2\sigma_y^2}\right)} \qquad (8)$$

has a normalized degree of coherence

$$\gamma(f_x, f_y) = e^{-\left(\frac{f_x^2}{2\sigma_{\gamma,x}^2} + \frac{f_y^2}{2\sigma_{\gamma,y}^2}\right)} \qquad (9)$$

The expression in the exponent of Eq. 9 indicates the coherence function $\gamma(f_x, f_y)$ is a continuum of concentric ellipses conjugate to the electron beam intensity profile.

Experimentally it is possible to rotate the double-slit field discriminator to measure the SR beam coherence as a function of observation angle θ. Relative to the x-axis, the projected spatial frequencies are then $f_x = \frac{d}{\lambda L}\cos\theta$ and $f_y = \frac{d}{\lambda L}\sin\theta$. For an upright Gaussian beam profile Eq. 8, the coherence function becomes

$$\gamma(f_x, f_y, \theta) = e^{-\left(\frac{\left(\frac{d}{\lambda L}\cos\theta\right)^2}{2\sigma_{\gamma,x}^2} + \frac{\left(\frac{d}{\lambda L}\sin\theta\right)^2}{2\sigma_{\gamma,y}^2}\right)} \qquad (10)$$

## EXPERIMENTAL RESULTS

The fringe contrast with the interferometer slits orientated in horizontal direction was first measured using a 530nm±5nm bandpass filter and σ-polarization of the SR beam. The raw data points and fitted Gaussian curve shown in Fig. 2 are plotted as a function of spatial frequency with an rms width of 1333 $m^{-1}$ (d=11.8 mm). From Eq. 6 we infer an rms beam size of $\sigma_x$=120 um. Taking into account the incoherent depth of field effect, the actual beam size is about 118 um [7].

Fringe contrast measured along the vertical axis is also plotted in Fig. 2. In this case the curve is wider due to the low aspect ratio of the electron beam. The rms width of the curve is 8135 m⁻¹ (d=72 mm) yielding an rms beam size of $\sigma_r$=20um. This value is well below the diffraction limit $\sigma_d$~80 um for direct visible SR imaging [8], and there is no incoherent depth of field effect. Both the horizontal and the vertical beam size measurements agree well with calculated values at the SR source.

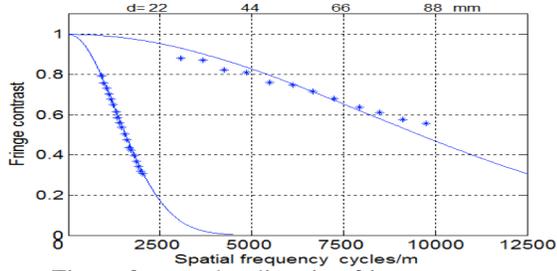

Figure 2: x- and y-direction fringe contrast.

Figure 3 shows the variation in fringe contrast when the double slit system and CCD camera were rotated synchronously around the SR beam axis and the polarizer angle held fixed. In this case the slits were separated by 16.94 mm and placed in the lower half of the SR beam below the cold finger. The double slit separation was a compromise between rms values of $\sigma_{\gamma,x}$~1333 m⁻¹ and $\sigma_{\gamma,y}$~8135 m⁻¹ taken from measurements seen in Fig. 2.

The four data sets plotted in Fig. 3 were measured by increasing betatron coupling leading to beam rotation angles of $\theta_0$=0°, 9°, 16° and 27°. The RMS sizes as measured along the minor axis of the beam ellipse were determined to be $\sigma_{minor}$=20, 35, 53 and 69 um. The theoretical curves according to Eq. 10 are superimposed.

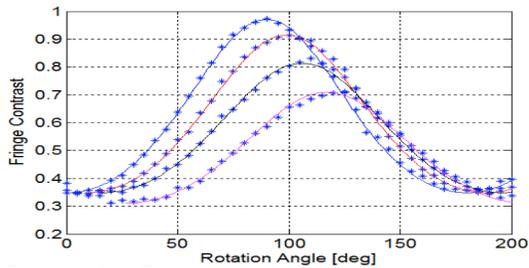

Figure 3: Contrast measured with double-slit rotation for 4 betatron coupling conditions.

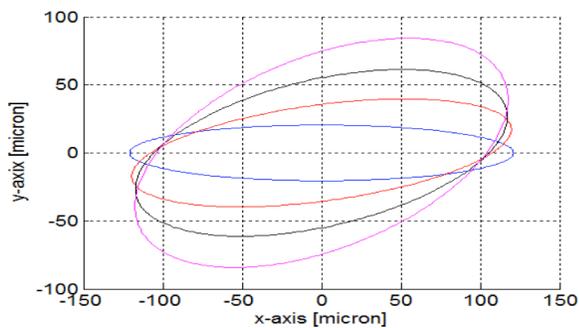

Figure 4: RMS electron beam cross-section color coded for each of the beam conditions in Fig. 3.

Figure 4 illustrates plots of the 2-D rms electron beam intensity profile derived from the fringe contrast data plotted in Fig. 3. The electron beam eigen-axes have been rotated by coupling of the betafunctions.

## CONCLUSION

In this paper we report the use of a Herschelian-type double-slit interferometer in use at the SPEAR3 visible light diagnostic beam line. The reflective geometry with commercial telescope eyepiece provides accurate, low aberration results. Measurements of the horizontal and vertical beam size agree well with the online accelerator model. Rotation of the double-slit mask with respect to the beam axis yields a modulated fringe contrast profile that can be used as a tomographic 'slicing' tool to extract the transverse electron beam intensity profile in agreement with theory.

## ACKNOWLEDGMENTS
The authors would like to thank the China Scholarship Council, the Fulbright Scholarship Council and members of the SPEAR3 operations team for support of this work.